# Web Single Sign-On Authentication using SAML

Kelly D. LEWIS, James E. LEWIS, Ph.D.

**Information Security, Brown-Forman Corporation**
Louisville, KY 40210, USA
*kellydlewis@gmail.com*

**Engineering Fundamentals, Speed School of Engineering, University of Louisville**
Louisville, KY 40292, USA
*jel@louisville.edu*

**Abstract**
Companies have increasingly turned to application service providers (ASPs) or Software as a Service (SaaS) vendors to offer specialized web-based services that will cut costs and provide specific and focused applications to users. The complexity of designing, installing, configuring, deploying, and supporting the system with internal resources can be eliminated with this type of methodology, providing great benefit to organizations. However, these models can present an authentication problem for corporations with a large number of external service providers. This paper describes the implementation of Security Assertion Markup Language (SAML) and its capabilities to provide secure single sign-on (SSO) solutions for externally hosted applications.
***Keywords:*** *Security, SAML, Single Sign-On, Web, Authentication*

## 1. Introduction

Organizations for the most part have recently started using a central authentication source for internal applications and web-based portals. This single source of authentication, when configured properly, provides strong security in the sense that users no longer keep passwords for different systems on sticky notes on monitors or under their keyboards. In addition, management and auditing of users becomes simplified with this central store.

As more web services are being hosted by external service providers, the sticky note problem has reoccurred for these outside applications. Users are now forced to remember passwords for HR benefits, travel agencies, expense processing, etc. - or programmers must develop custom SSO code for each site. Management of users becomes a complex problem for the help desk and custom built code for each external service provider can become difficult to administer and maintain.

In addition, there are problems for the external service provider as well. Every user in an organization will need to be set up for the service provider's application, causing a duplicate set of data. Instead, if the organization can control this user data, it would save the service provider time by not needing to set up and terminate user access on a daily basis. Furthermore, one central source would allow the data to be more accurate and up-to-date.

Given this set of problems for organizations and their service providers, it is apparent that a solution is needed that provides a standard for authentication information to be exchanged over the Internet. Security Assertion Markup Language (SAML) provides a secure, XML-based solution for exchanging user security information between an identity provider (our organization) and a service provider (ASPs or SaaSs). The SAML standard defines rules and syntax for the data exchange, yet is flexible and can allow for custom data to be transmitted to the external service provider.

## 2. Background

The consortium for defining SAML standards and security is OASIS (Organization for the Advancement of Structured Information Standards). They are a non-profit international organization that promotes the development and adoption of open standards for security and web services. OASIS was founded in 1993 under SGML (Standard Generalized Markup Language) Open until its name change in 1998. Headquarters for OASIS are located in North America, but there is active member participation internationally in 100 countries on five continents [1].

IJCSI

SAML 1.0 became an OASIS standard toward the end of 2002, with its early formations beginning in 2001. The goal behind SAML 1.0 was to form a XML framework to allow for the authentication and authorization from a single sign-on perspective. At the time of this milestone, other companies and consortiums started extending SAML 1.0. While these extensions were being formed, the SAML 1.1 specification was ratified as an OASIS standard in the fall of 2003.

The next major revision of SAML is 2.0, and it became an official OASIS Standard in 2005. SAML 2.0 involves major changes to the SAML specifications. This is the first revision of the standard that is not backwards compatible, and it provides significant additional functionality [2]. SAML 2.0 now supports W3C XML encryption to satisfy privacy requirements [3]. Another advantage that SAML 2.0 includes is the support for service provider initiated web single sign-on exchanges. This allows for the service provider to query the identity provider for authentication. Additionally, SAML 2.0 adds "Single Logout" functionality. The remainder of this text will be discussing implementation of a SAML 2.0 environment.

There are three roles involved in a SAML transaction – an asserting party, a relying party, and a subject. The asserting party (identity provider) is the system in authority that provides the user information. The relying party (service provider) is the system that trusts the asserting party's information, and uses the data to provide an application to the user. The user and their identity that is involved in the transaction are known as the subject.

The components that make up the SAML standard are assertions, protocols, bindings and profiles. Each layer of the standard can be customized, allowing specific business cases to be addressed per company. Since each company's scenarios could be unique, the implementation of these business cases should be able to be personalized per service and per identity providers.

The transaction from the asserting party to the relying party is called a SAML assertion. The relying party assumes that all data contained in the assertion from the asserting party is valid. The structure of the SAML assertion is defined by the XML schema and contains header information, the subject and statements about the subject in the form of attributes and conditions. The assertion can also contain authorization statements defining what the user is permitted to do inside the web application.

The SAML standard defines request and response protocols used to communicate the assertions between the service provider (relying party) and the identity provider (asserting party). Some example protocols are [4]:

- Authentication Request Protocol – defines how the service provider can request an assertion that contains authentication or attribute statements
- Single Logout Protocol – defines the mechanism to allow for logout of all service providers
- Artifact Resolution Protocol – defines how the initial artifact value and then the request/response values are passed between the identity provider and the service provider.
- Name Identifier Management Protocol – defines how to add, change or delete the value of the name identifier for the service provider

SAML bindings map the SAML protocols onto standard lower level network communication protocols used to transport the SAML assertions between the identity provider and service provider. Some example bindings used are [4]:

- HTTP Redirect Binding – uses HTTP redirect messages
- HTTP POST Binding – defines how assertions can be transported using base64-encoded content
- HTTP Artifact Binding – defines how an artifact is transported to the receiver using HTTP
- SOAP HTTP Binding – uses SOAP 1.1 messages and SOAP over HTTP

The highest SAML component level is profiles, or the business use cases between the service provider and the identity provider that dictate how the assertion, protocol and bindings will work together to provide SSO. Some example profiles are [4]:

- Web Browser SSO Profile – uses the Authentication Request Protocol, and any of the following bindings: HTTP Redirect, HTTP POST and HTTP Artifact
- Single Logout Profile – uses the Single Logout Protocol, which can log the user out of all service providers using a single logout function
- Artifact Resolution Profile – uses the Artifact Resolution Protocol over a SOAP HTTP binding
- Name Identifier Management Profile – uses the name Identifier management Protocol and can be used with HTTP Redirect, HTTP POST, HTTP Artifact or SOAP

Two profiles will be briefly discussed in more detail, the artifact resolution profile and web browser SSO profile. The artifact resolution profile can be used if the business case requires highly sensitive data to pass between the identity provider and service provider, or if the two partners want to utilize an existing secure connection between the two companies.

This profile allows for a small value, called an artifact to be passed between the browser and the service provider by one of the HTTP bindings. After the service provider receives the artifact, it transmits the artifact and the request/response messages out of band from the browser back to the identity provider. Most likely the messages are transmitted over a SSL VPN connection between the two companies. This provides security for the message, plus eliminates the need for the assertions to be signed or encrypted which could potentially reduce overhead. When the identify provider receives the artifact, it looks up the value in its database and processes the request. After all out of band messages are transmitted between the identity provider and service provider, the service provider presents the information directly to the browser.

The web browser SSO profile may be initiated by the identify provider or the service provider. If initiated by the identity provider, the assertion is either signed, encrypted, or both. In the web browser SSO profile, all of the assertion information is sent at once to the service provider using any of the HTTP bindings and protocols. The service provider decrypts if necessary and checks for message integrity against the signature. Next, it parses the SAML XML statements and gathers any attributes that were passed, and then performs SSO using the Assertion Consumer Service. The diagram in Figure 1 shows the identity provider initiated SAML assertion.

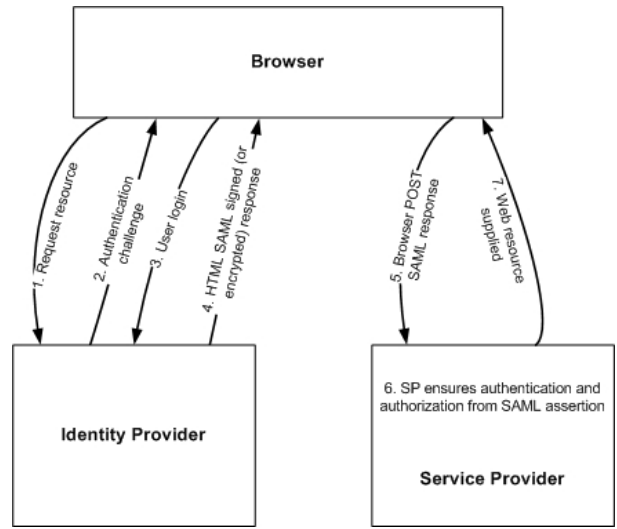

Figure 1: Identity Provider Initiated SAML Assertion Flowchart

If the user accesses the external webpage without passing through the internal federated identity manager first, the service provider will need to issue the SAML request back to the identity provider on behalf of the user. This process of SSO is called service provider initiated. In this case, the user arrives at a webpage specific for the company, but without a SAML assertion. The service provider redirects the user back to the identity provider's federation webpage with a SAML request, and optionally with a RelayState query string variable that can be used to determine what SAML entity to utilize when sending the assertion back to the service provider.

After receiving the request from the service provider, the identity provider processes the SAML request as if it came internally. This use case is important since it allows users to be able to bookmark external sites directly, but still provides SAML SSO capabilities with browser redirects. Figure 2 demonstrates this service provider initiated use case.

The most popular business use case for SAML federation is the web browser SSO profile, used in conjunction with the HTTP POST binding and authentication request protocol. The implementation and framework section will discuss this specific use case and the security needed to protect data integrity.

IJCSI

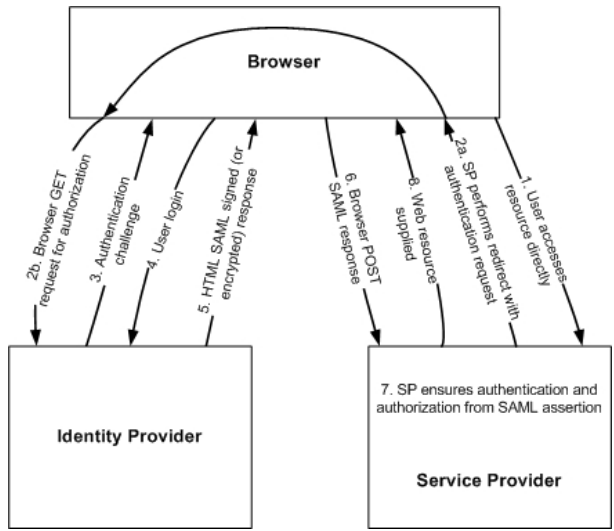

Figure 2: Service Provider Initiated SAML Assertion Flowchart

## 3. Implementation/Framework

There are numerous identity and federation manager products on the market that support federation via SAML versions 1.1 and 2.0, as well as several open source products. OpenSAML, an open source toolkit, is available to support developers working with SAML. Shibboleth is an example of an open source project that uses the OpenSAML toolkit. Sun Microsystems has a product called OpenSSO that is an open source version of their commercial product, OpenSSO Enterprise. Computer Associates provides an access manager called SiteMinder and RSA has a product called Federated Identity Manager to name a few. Regardless of which product is selected, as long as it conforms to the standards of SAML, all products can be used interchangeably with no compatibility issues.

The process of setting up federation involves configuring a local entity, a partner entity, and an association between the two that forms the federation. The local entity must be manually configured inside the federation software; however, for SAML 2.0 the process of setting up the partner entity has been made easier with the introduction of metadata. Since the SAML standard is flexible and can allow a number of custom configurations, certain agreements and configuration information must be initially set up between two partners. Exchanging metadata containing this specific information determines the specifications that will be used in a particular business case.

Once the metadata file has been received from the partner entity, this XML file can be uploaded into the federation software without any additional configuration needed for the partner entity. This process saves time and reduces the possibility for error. The file contains elements and attributes, and includes an EntityDescriptor and EntityID that specifies to which entity the configuration refers.

There are many optional elements and attributes for metadata files; some that may apply are Binding, WantAuthRequestsSigned, WantAssertionsSigned, SingleLogoutService, etc. To review the entire list of elements available for the metadata file, see the OASIS metadata standard [5].

When manually configuring a local entity, first determine the parameters to be passed in the assertion that will be the unique username for each user. Normally this value is an email address or employee number, since they are guaranteed to be exclusive for each individual. In some federation products, values from a data source can be automatically utilized with the SAML assertion. These values can be extracted from different data sources such as LDAP, or another source that could be tied into a HR system. While setting up the local entity there are other considerations, such as how the parameters will be passed (in attributes or nameID), a certificate keystore for the association, and type of signing policies required.

The following sample metadata shown in Figure 3 is an example that would be sent from the local entity (identity provider in this case) to the partner entity (service provider) to load into the federation software. The descriptor shows titled as "IDPSSODescriptor", which demonstrates this is metadata from an identity provider.

Some elements are mandatory, such as entityID, yet others are optional, such as ID and OrganizationName. The elements to note are the Single Sign-On Service binding, location, protocol support section, and key descriptor and key info areas. In this example, the binding must be performed by an HTTP-POST, and the supported protocol is SAML 2.0.

IJCSI

```xml
<md:EntityDescriptor ID="MyCompany"
  entityID="mycompany:saml2.0"
  xmlns:ds="http://www.w3.org/2000/09/xmldsig#"
  xmlns:md="urn:oasis:names:tc:SAML:2.0:metadata"
  xmlns:query="urn:oasis:names:tc:SAML:metadata:ext:query"
  xmlns:saml="urn:oasis:names:tc:SAML:2.0:assertion"
  xmlns:xenc="http://www.w3.org/2001/04/xmlenc#">
 <md:IDPSSODescriptor WantAuthnRequestsSigned="false"
  protocolSupportEnumeration=
    "urn:oasis:names:tc:SAML:2.0:protocol">
  <md:KeyDescriptor use="encryption">
   <ds:KeyInfo
    xmlns:ds="http://www.w3.org/2000/09/xmldsig#">
      <ds:X509Data>
       <ds:X509Certificate>
        CERTIFICATE
       </ds:X509Certificate>
      </ds:X509Data>
   </ds:KeyInfo>
   <md:EncryptionMethod
    Algorithm="http://www.w3.org/2001/04/xmlenc#aes128-cbc">
   </md:EncryptionMethod>
  </md:KeyDescriptor>
  <md:KeyDescriptor use="signing">
   <ds:KeyInfo xmlns:ds="http://www.w3.org/2000/09/xmldsig#">
      <ds:X509Data>
       <ds:X509Certificate>
        CERTIFICATE
       </ds:X509Certificate>
      </ds:X509Data>
   </ds:KeyInfo>
  </md:KeyDescriptor>
  <md:SingleSignOnService
    Binding="urn:oasis:names:tc:SAML:2.0:bindings:HTTP-POST"
    Location="http://mycompany.com/sso/SSO">
  </md:SingleSignOnService>
 </md:IDPSSODescriptor>
 <md:Organization>
  <md:OrganizationName xml:lang="en-us">
   My Company Org
  </md:OrganizationName>
  <md:OrganizationDisplayName xml:lang="en-us">
   My Company
  </md:OrganizationDisplayName>
  <md:OrganizationURL xml:lang="en-s">
   http://www.mycompany.com
  </md:OrganizationURL>
 </md:Organization>
</md:EntityDescriptor>
```
Figure 3: Sample Identity Provider Metadata XML

Figure 4 demonstrates an example metadata XML file that would be sent from a service provider to an identity provider for loading into the federation software. Note that the descriptor is "SPSSODescriptor", indicating service provider single sign-on descriptor.

In this case, "WantAuthnRequestsSigned" is equal to true, as opposed to the previous example in Figure 3.

Also, there are two KeyDescriptors, one for signing and one for encrypting. This indicates the service provider requires both for the assertion. There are two methods of binding listed for the assertion consumer service: the HTTP Post and the HTTP Artifact. These two metadata samples show how custom each company can be with unique SAML requirements.

```xml
<EntityDescriptor
  entityID="mypartner:saml2.0"
  xmlns="urn:oasis:names:tc:SAML:2.0:metadata">
 <SPSSODescriptor
   AuthnRequestsSigned="true"
   WantAssertionsSigned="true"
   protocolSupportEnumeration=
     "urn:oasis:names:tc:SAML:2.0:protocol">
  <KeyDescriptor use="signing">
   <ds:KeyInfo xmlns:ds="http://www.w3.org/2000/09/xmldsig#">
    <ds:X509Data>
     <ds:X509Certificate>CERTIFICATE</ds:X509Certificate>
    </ds:X509Data>
   </ds:KeyInfo>
  </KeyDescriptor>
  <KeyDescriptor use="encryption">
   <ds:KeyInfo xmlns:ds="http://www.w3.org/2000/09/xmldsig#">
    <ds:X509Data>
     <ds:X509Certificate>CERTIFICATE</ds:X509Certificate>
    </ds:X509Data>
   </ds:KeyInfo>
   <EncryptionMethod
    Algorithm="http://www.w3.org/2001/04/xmlenc#aes128-cbc">
    <xenc:KeySize
     xmlns:xenc="http://www.w3.org/2001/04/xmlenc#">128
    </xenc:KeySize>
   </EncryptionMethod>
  </KeyDescriptor>
  <NameIDFormat>
   urn:oasis:names:tc:SAML:2.0:nameid-format:transient
  </NameIDFormat>
  <AssertionConsumerService
   index="0"
   isDefault="true"
   Binding="urn:oasis:names:tc:SAML:2.0:bindings:HTTP-POST"
   Location="https://mypartner.com/federation/metaAlias/sp"/>
  <AssertionConsumerService
   index="1"
   Binding="urn:oasis:names:tc:SAML:2.0:bindings:HTTP-Artifact"
   Location="https://mypartner.com/federation/metaAlias/sp"/>
 </SPSSODescriptor>
</EntityDescriptor>
```
Figure 4: Sample Service Provider Metadata XML

After the metadata is exchanged and all entities are set up, the assertion can be tested and verified using browser tools and decoders. For this example, the service provider implementation of the HTTP POST method will be described briefly.

The identity provider must first determine what URL the federation software requires, and what attributes need to be passed with the POST data, such as entityID or RelayState. The browser HTTP-POST action contains hidden SAMLResponse and RelayState fields enclosed in a HTML form. After the browser POST is received by the service provider, the Assertion Consumer Service validates the signature and processes the assertion, gathering attributes and other conditions that could optionally be required. The service provider also obtains the optional RelayState variable in the HTML form, determines the application URL, and redirects the browser to it providing single sign-on to the web application [4].

To validate the sent attributes in the assertion with this HTTP POST example, a browser add-on program can be used to watch exactly what is sent between the browser and the partner. A few browser add-ons are "HttpFox" [6] which can be used with Mozilla Firefox, and "HttpWatch" [7] which can be used with Mozilla Firefox or Internet Explorer. After capturing HTTP data, the browser POST action can be verified to ensure the proper attributes are passed to the partner. The POST action shows the hidden SAMLResponse and RelayState fields in the HTML form, and can be used to validate the data sent to the service provider.

The SAMLResponse field is URL encoded, and must be decoded before reading the assertion. Depending on the requirements, the assertion must be signed, or signed and encrypted. For testing purposes, first only sign the assertion so it can be URL decoded into a non-encrypted readable version. Figure 5 shows an example of a URL decoded SAMLResponse and has been shortened for readability, designated by capital words.

```
<samlp:Response xmlns:saml="urn:oasis:names:tc:SAML:2.0:assertion"
        xmlns:samlp="urn:oasis:names:tc:SAML:2.0:protocol"
        Consent="urn:oasis:names:tc:SAML:2.0:consent:unspecified"
        Destination="https://mypartner.com/metaAlias/sp"
        ID="ad58514ea9365e51c382218fea"
        IssueInstant="2009-04-22T12:33:36Z"
        Version="2.0">
 <saml:Issuer>http://login.mycompany.com/mypartner</saml:Issuer>
 <ds:Signature xmlns:ds="http://www.w3.org/2000/09/xmldsig#">
   SIGNATURE VALUE, ALGORITHM, ETC.
 </ds:Signature>
 <samlp:Status>
  <samlp:StatusCode
        Value="urn:oasis:names:tc:SAML:2.0:status:Success">
  </samlp:StatusCode>
 </samlp:Status>
 <saml:Assertion ID="1234" IssueInstant="2009-04-22T12:33:36Z"
        Version="2.0">
  <saml:Issuer>http://login.mycompany.com/mypartner</saml:Issuer>
  <ds:Signature xmlns:ds="http://www.w3.org/2000/09/xmldsig#">
   SIGNATURE VALUE, ALGORITHM, ETC.
  </ds:Signature>
  <saml:Subject>
   <saml:NameID>NAMEID FORMAT, INFO, ETC</saml:NameID>
   <saml:SubjectConfirmation
        Method="urn:oasis:names:tc:SAML:2.0:cm:bearer">
    <saml:SubjectConfirmationData
        NotOnOrAfter="2009-04-22T12:43:36Z"
        Recipient="https://mypartner.com/metaAlias/sp">
    </saml:SubjectConfirmationData>
   </saml:SubjectConfirmation>
  </saml:Subject>
  <saml:Conditions
        NotBefore="2009-04-22T12:28:36Z"
        NotOnOrAfter="2009-04-22T12:33:36Z">
   <saml:AudienceRestriction>
    <saml:Audience>mypartner.com:saml2.0</saml:Audience>
   </saml:AudienceRestriction>
  </saml:Conditions>
  <saml:AuthnStatement AuthnInstant="2009-04-22T12:33:20Z"
        SessionIndex="ccda16bc322adf4f74d556bd">
   <saml:SubjectLocality Address="192.168.0.189"
        DNSName="myserver.mycompany.com">
   </saml:SubjectLocality>
  </saml:AuthnStatement>
  <saml:AttributeStatement xmlns:xs=SCHEMA INFO>
   <saml:Attribute FriendlyName="clientId" Name="clientId"
        NameFormat="urn:oasis:names:tc:SAML:2.0:
           attrname-format:basic">
    <saml:AttributeValue>1234</saml:AttributeValue>
   </saml:Attribute>
   <saml:Attribute FriendlyName="uid" Name="uid"
        NameFormat="urn:oasis:names:tc:SAML:2.0:
           attrname-format:basic">
    <saml:AttributeValue>the.user@mycompany.com
    </saml:AttributeValue>
   </saml:Attribute>
  </saml:AttributeStatement>
 </saml:Assertion>
</samlp:Response>
```

Figure 5:  Sample SAML Assertion

For testing purposes with this sample assertion, the attributes toward the end of the XML should be verified. In this example, two attributes are being passed: clientID and uid. The clientID is a unique value that has been assigned by the service provider indicating which company is sending the assertion. The uid in this case is the email address of the user requesting the web resource. After receiving and validating these values, the service provider application performs SSO for the user. Once these values have been tested and accepted as accurate, the SAML assertion can be encrypted if required, and the service provider application can be fully tested.

There are important security aspects to be considered, given that the relying party fully trusts the data in the SAML assertion. The integrity of the message must be preserved from man-in-the-middle attacks and other spoofs. In dealing with this scenario, A SAML assertion can be unsigned, signed, or signed and encrypted depending on the type of data and the sensitivity required per application. The SAML standard allows for message integrity by supporting X509 digital signatures in the request/response transmissions. SAML also supports and recommends HTTP over SSL 3.0 and TLS 1.0 for situations where data confidentiality is required [8].

As analyzed by Hansen, Skriver, and Nielson there are some major issues in the SAML 1.1 browser/artifact profile using TLS security [9]. In SAML 2.0, this profile was improved to repair a majority of these security issues; however there is one existing problem in the specification examined by Groß and Pfitzmann [10]. Groß and Pfitzmann devised a solution to this exploit by creating a new profile that produces two artifacts, with the token being valid only when it consists of both values, thus eliminating successful replay of a single token. Additional work has also been performed on recently proposed attack scenarios. Gajek, Liao, and Schwenk recommend two new stronger bindings for SAML artifacts to the TLS security layer [11].

An additional scenario that could compromise data integrity is a replay attack that intercepts the valid assertion and saves the data for impersonation at a later time. Both the identity provider and the service provider should utilize the SAML attributes NotBefore and NotOnOrAfter shown in Figure 5. These values should contain a time frame that is as short as possible, usually around 5 minutes. In addition, the identity provider can insert locality information into the assertion, which the service provider can verify is valid against the IP address of the requesting user. For additional security considerations, see the OASIS security and privacy considerations standard [8].

## 4. Conclusions/Best Practices

In conclusion, the benefits of SAML are abundant. Organizations can easily, yet securely share identity information and security is improved by eliminating the possibility of shared accounts. User experience is enhanced by eliminating additional usernames and passwords, which also allows for fewer helpdesk calls and administrative costs.

Companies should have documentation available to exchange when setting up SAML associations, since each SAML use case can be customized per individual business need. Service providers can use different security protocols, such as signed only, versus signed and encrypted. In addition, some service providers may only use the nameID section of the assertion, while others might use custom attributes only. This upfront documentation can save troubleshooting time during the implementation and testing phases of the project.

Furthermore, during testing phases it is helpful to use a sample test site for the service provider and also to test with SAML assertions signed only. The sample test site allows for the ability to isolate a test of only the SAML connection between the two partners, before testing of the application occurs. Testing with signed only assertions allows for the ability to URL decode the HTML hidden input field, and validate the data being passed to the service provider. This ensures the correct data in the assertion is sent and can be tested prior to the service provider site being fully prepared for testing.

Additionally, using SAML metadata is very helpful since it eliminates typos and errors when setting up the partner entity. These metadata files can help the identity provider understand exactly what the service provider needs in the SAML assertion. Both the identity provider and service provider should utilize metadata files, not only to speed up manual work when entering data into the federation software, but to also reduce human error.

The OASIS Security Services Technical Committee continues to improve upon the current SAML 2.0 standard by developing new profiles to possibly be used in later releases. For example, one area OASIS has already improved upon was a supplement to the metadata specifications that added new elements and descriptor types. Both identity providers and service providers should be aware of any changes to SAML standards that are ratified by OASIS. Staying current and not deviating from the standards helps to ensure compatibility, resulting in less customized configurations between organizations.

IJCSI

**Kelly D. Lewis** graduated with a B.S. of Computer Engineering and Computer Science at the University of Louisville in 2001. She received her M. Eng. at the University of Louisville in the same discipline in 2005, publishing a thesis titled "Student Performance Evaluation Package using a Web Interface and a Database". She started her Information Technology career in 1999 with the United States Army Research Institute. She has been employed for Brown-Forman Corporation the last 8 years, and has worked a Systems Administrator, Network Engineer, and presently holds a Security Analyst position in Information Security. Her focus is on network security, automation, and single sign-on technologies.

**James E. Lewis** graduated with a B.A. in Computer Science from Hanover College in 1994, and earned a M.S. in Computer Science from the University of Louisville in 1996, with a thesis focusing on expert systems and networking. He received a Ph.D. in Computer Science and Engineering from the University of Louisville in 2003, publishing a dissertation with an emphasis in distributed genetic algorithms. He started teaching in 1995, and is currently an Assistant Professor in the Department of Engineering Fundamentals at the University of Louisville's Speed School of Engineering, where he received this appointment in 2004. He has fourteen publications on various topics including distributed algorithms, intelligent system design, and engineering education that were published in national and international conference proceedings. He has also been invited to present on critical thinking in engineering education at two conferences. He has been awarded two research grants for his critical thinking and case study initiatives. He is a member of the ACM and ASEE organizations. His research interests include parallel and distributed computer systems, cryptography, security design, engineering education, and technology used in the classroom.